\documentclass[12pt]{article}%
\usepackage{amsfonts}
\usepackage{sw20elba}%
\usepackage{amsmath}%
\usepackage{amssymb}%
\usepackage{graphicx}
\providecommand{\U}[1]{\protect\rule{.1in}{.1in}}
\begin{document}

\title{Anomalous He-Gas High-Pressure Studies on Superconducting LaO$_{1-x}$F$_{x}$FeAs}
\author{W. Bi, H. B. Banks and J. S. Schilling\\Department of Physics, Washington University\\CB 1105, One Brookings Dr, St. Louis, MO 63130, USA
\and H. Takahashi and H. Okada\\Department of Physics, College of Humanities \& Sciences\\Nihon University, Sakurajousui, Setagaya-ku, Tokyo 156-8550, Japan
\and Y. Kamihara, M. Hirano and H. Hosono\\Frontier Research Center, Tokyo Institute of Technology\\4259 Nagatsuda, Midori-ku, Yokohama 226-8503, Japan}
\date{November 26, 2009 }
\maketitle

\begin{abstract}
AC susceptibility measurements have been carried out on superconducting
LaO$_{1-x}$F$_{x}$FeAs for $x=0.07$ and $0.14$ under He-gas pressures to
$\sim$ 0.8 GPa. Not only do the measured values of $dT_{c}/dP$ differ
substantially from those obtained in previous studies using other pressure
media, but the $T_{c}(P)$ dependences observed depend on the detailed
pressure/temperature history of the sample. A sizeable sensitivity of
$T_{c}(P)$ to shear stresses provides a possible explanation.

\end{abstract}

\newpage

\section{Introduction}

The discovery of superconductivity at temperatures as high as 26 K in
LaO$_{1-x}$F$_{x}$FeAs \cite{kamihara}, a layered compound devoid of CuO$_{2}$
planes, has rekindled interest in high-temperature superconductivity. As for
the cuprates \cite{schilling1}, high pressure experiments can potentialy play
an important role in furthering our understanding of the new Fe-based
pnictides \cite{takahashi1,chu}. High-pressure experiments to 30 GPa by
Takahashi \textit{et al.} \cite{takahashi2} on this compound for $x=0.11$
using a solid (NaCl) pressure medium reveal that the superconducting onset in
the electrical resistivity reaches temperatures as high as 46 K. The initial
slope $dT_{c}/dP$ appears to increase with the fluorine content, as least for
concentrations $x\leq0.14$ \cite{takahashi1}. Resistivity and magnetic
susceptibility studies by several groups \cite{takahashi1,takahashi2,lu,zocco}
using various fluid pressure media agree that $dT_{c}/dP$ is positive
initially, but differ widely on its magnitude. As in previous studies on the
high-$T_{c}$ cuprates \cite{schilling1} and the binary compound MgB$_{2}$
\cite{deemyad}, it would be of interest to carry out benchmark determinations
of $T_{c}(P)$ in magnetic susceptibility measurements on LaO$_{1-x}$F$_{x}%
$FeAs using the most hydrostatic pressure medium known, He gas.

One notable result is that the non-superconducting, undoped pnictides LaOFeAs
\cite{takahashi1}, CaFe$_{2}$As$_{2}$ \cite{tori}, SrFe$_{2}$As$_{2}$
\cite{alireza}, and BaFe$_{2}$As$_{2}$ \cite{alireza} reportedly become
superconducting under pressure when pressure media such as Fluorinert,
methanol-ethanol, and silicone oil were used, superconductivity in CaFe$_{2}%
$As$_{2}$ appearing at a relatively low pressure $\leq0.4$ GPa. However,
recent dc susceptibility and electrical resistivity measurements by Yu
\textit{et al.} \cite{yu} on CaFe$_{2}$As$_{2}$ using He-gas pressure medium
fail to detect any sign of superconductivity to 0.7 GPa and 2 K. These authors
suggest that not only in CaFe$_{2}$As$_{2},$ but also in SrFe$_{2}$As$_{2}$
and BaFe$_{2}$As$_{2}$, non-hydrostatic stress components may have been
responsible for the reported pressure-induced superconductivity. In addition,
very recently Matsubayashi \textit{et al.} \cite{mat} carried out parallel ac
susceptibility and electrical resistivity measurements on BaFe$_{2}$As$_{2}$
and SrFe$_{2}$As$_{2}$ single crystals to 8 GPa pressure in a relatively
hydrostatic cubic anvil pressure apparatus and find no trace of
superconductivity in the former compound. In SrFe$_{2}$As$_{2}$ bulk
superconductivity, as evidenced by full shielding in the ac susceptibility, is
only found in a rather narrow (2 GPa) pressure region centered about 6 GPa,
whereas non-bulk (filamentary) superconductivity is revealed in the
resistivity which falls to zero over a much wider pressure region. The
potentially important role shear stresses play in the superconductivity of the
Fe-based pnictides is emphasized in the recent review of Chu and Lorenz
\cite{chu}. Indeed, shear-stress effects on $T_{c}(P)$ are well known from
studies on such diverse superconducting materials as organic metals
\cite{schirber2}, high-$T_{c}$ oxides \cite{schilling1}, MgB$_{2}$
\cite{deemyad}, and Re metal \cite{chu2}.

To our knowledge, the measurements of Yu \textit{et al.} \cite{yu} on
CaFe$_{2}$As$_{2}$ are the only He-gas high-pressure studies of
superconductivity yet carried out on an Fe-based pnictide. It would be of
obvious interest to extend such studies to the LaO$_{1-x}$F$_{x}$FeAs system
to ascertain whether $T_{c}(P)$ differs from the findings of earlier studies
where less hydrostatic pressure media were employed. Using He as pressure
medium brings a further benefit: \ it allows one to change the hydrostatic
pressure at relatively low temperatures, thus permitting studies of phenomena
impacting superconductivity which are both pressure and temperature dependent.
An example for such phenomena in the cuprates are the well known oxygen
ordering effects which have been shown in some systems to be the dominant
factor determining the dependence of $T_{c}$ on pressure \cite{schilling1}.

To throw some light on these issues, we have determined $T_{c}(P)$ using
He-gas pressure medium to 0.78 GPa for the original Fe-based superconducting
pnictide, LaO$_{1-x}$F$_{x}$FeAs, where $x=0.07$ and 0.14. For these doping
levels we find the initial pressure derivative $dT_{c}/dP$ to be positive, but
markedly less than the published values using other pressure media; in
addition, the dependence of $T_{c}$ on pressure is not reversible but depends
on the detailed pressure/temperature history of the sample. Possible origins
for this behavior are discussed.

\section{Experiment}

Polycrystalline LaO$_{1-x}$F$_{x}$FeAs samples are prepared by solid-state
reaction as described in previous publications\textit{\ }%
\cite{kamihara,takahashi2}. For $x=0.14$ both the previous resistivity
measurements \cite{takahashi1} and the present ac susceptibility studies are
carried out on pieces taken from the same mother sample. The fluorine content
in the present samples, $x=0.07$ and 0.14, is determined from the lattice
constant using Vegard's law; these samples have densities 6.704 g/cm$^{3}$ and
6.739 g/cm$^{3}$ \cite{kamihara} and masses 1.54 mg and 8.69 mg (or 5.65 mg), respectively.

To generate hydrostatic pressures as high as 0.8 GPa a He-gas compressor
system (Harwood Engineering) was used in combination with a CuBe pressure cell
(Unipress). AC susceptibility measurements at 0.1 Oe rms and 1023 Hz were
carried out under pressure to the same high accuracy as measurements at
ambient pressure by surrounding the sample with a primary/secondary
compensated coil system connected to a Stanford Research SR830 digital lock-in
amplifier via a SR554 transformer preamplifierthe. A two-stage closed-cycle
refrigerator was used to cool the pressure cell to temperatures as low as 5-6
K; measurements were carried out upon warming up slowly through the transition
at the rate $\sim0.3$ K/min. All susceptibility measurements were repeated at
least once to verify that the reproducibility of the transition temperature
was within 20 mK. Unless otherwise stated, the sample was cooled down to
measure $T_{c}$ 30-60 min after a given change in pressure/temperature.
Further details of the He-gas techniques are given elsewhere\textit{\ }%
\cite{techniques}.

\section{Experimental Results}

\subsection{LaO$_{0.93}$F$_{0.07}$FeAs sample}

For the $x=0.07$ sample at ambient pressure the 80-20 transition width in the
ac susceptibility is $\sim$ 1 K where $T_{c}\simeq21.7$ K from the transition
onset and $T_{c}\simeq21$ K from the midpoint (see Fig.~1(a)). These values of
$T_{c}$ are several degrees Kelvin less than those estimated from the
resistivity onset at comparable fluorine concentrations \cite{takahashi1}.

Previous high-pressure resistivity measurements with Fluorinert pressure
medium on LaFeAsO$_{1-x}$F$_{x}$ for $x=0.05$, 0.08, and 0.11 yielded the
following positive values of the initial derivative $dT_{c}/dP\simeq$ 2, 2,
and 8 K/GPa, respectively, where $T_{c}$ was determined from the resistivity
onset \cite{takahashi1}. From these results one would anticipate that
$dT_{c}/dP\simeq2$ K/GPa for $x=0.07$. On the other hand, if $T_{c}$ is
determined from the temperature at which the resistivity $\rho$ drops to 0,
one finds for $x=0.08$ the initial pressure dependence $dT_{c}/dP\simeq0.54 $
K/GPa, a value approximately 4$\times$ smaller than that from the resistivity
onset (for $x=0.05$ and 0.11 it was not possible to reliably estimate the
temperature at which $\rho\rightarrow0).$ In resistivity measurements,
therefore, the value of $dT_{c}/dP$ obtained evidently depends on the
criterion used to determine $T_{c}$.

We now compare the values of $dT_{c}/dP$ obtained from the present ac
susceptibility measurements using hydrostatic He-gas pressure on
LaFeAsO$_{1-x}$F$_{x}$ to those obtained in the above resistivity studies. In
Fig.~1(a) are shown our results for the F concentration $x=0.07$. The large
magnitude of the superconducting transition is consistent with bulk
superconductivity; in fact, the shielding effect is approximately twice that
expected for perfect diamagnetism. No correction is made here for
diamgnetization effects. Under increasing He-gas pressure to 0.78 GPa, the
superconducting transition is seen to shift monotonically to higher
temperatures. Here $T_{c}$ is defined by the transition midpoint (see
Fig.~1(a)); however, since the shape $\chi^{\prime}(T)$ of the transition does
not change with pressure, the shift in $T_{c}$ with pressure is the same
whether $T_{c}$ is defined from the transition midpoint or onset.

In Fig.~1(b) the dependence of $T_{c}$ on pressure is shown for all
measurements on the $x=0.07$ sample, the numbers giving the order of
measurement. After the ambient pressure measurement (point 1), 0.78 GPa
pressure is applied at room temperature (RT) to yield point 2. $T_{c}$ is seen
to increase under pressure at the rate +1.20 K/GPa, clearly less than the
value +2 K/GPa inferred from the resistivity onset but greater than the value
using the $\rho\rightarrow0$ criterion at nearly the same F concentration
($x=0.08$) \cite{takahashi1}; in the resistivity studies the pressure was
always changed at RT. The pressure was then successively reduced at low
temperatures (62 K, 52 K, 45 K for points 2$\rightarrow$3, 3$\rightarrow$4,
and 4$\rightarrow$5, respectively) before cooling down further to measure
$T_{c}.$ Up to and including point 4, the $T_{c}(P)$ dependence is reversible;
however, at ambient pressure (point 5) $T_{c}$ lies $\sim220$ mK lower than
the initial value at ambient pressure (point 1). Interestingly, after warming
the sample back to RT and holding for 1.5 h, $T_{c}$ is seen to revert (point
6) back to its initial value. This behavior bears some resemblence to that
observed previously from oxygen-ordering effects in the cuprates where
$T_{c}(P)$ may differ strongly whether the pressure is changed at RT or low
temperatures \cite{schilling1}; there is, however, one notable difference - -
when pressure is reduced at low temperature in a cuprate with oxygen ordering,
all values of $T_{c}$ would differ, and not just those below a certain
pressure threshold.

To examine whether, as in the cuprates, there exists a particular (sub-RT)
temperature above which such $T_{c}$-relaxation occurs, we applied 0.66 GPa
pressure at RT (point 7) and then released pressure at 60 K (point 8),
reproducing exactly the previous results. Holding the sample at 100 K for 90
min resulted in no change in $T_{c}$ (point 9). However, after holding the
sample at 200 K for 100 min, $T_{c}$ returned to its initial value (point 10).
Warming back to RT for 1 h (point 11) and then for one week (point 12),
resulted in no further change in $T_{c}.$ A pressure of 0.51 GPa was then
applied at 60 K (point 13) and released at 50 K (point 14); these $T_{c}$
values faithfully track the +1.2 K/GPa straight line in Fig.~1(b).

\subsection{LaO$_{0.86}$F$_{0.14}$FeAs sample}

For the $x=0.14$ sample at ambient pressure the 80-20 transition width in the
ac susceptibility is $\sim$ 2 K where $T_{c}\simeq13.7$ K from the transition
onset and $T_{c}\simeq12.5$ K from the midpoint (see Fig.~2(a)). This onset
value is several degrees Kelvin less than from the resistivity onset ($\sim19$
K); however, the temperature of the susceptibility midpoint is comparable to
the resistivity zero point ($\sim12$ K) \cite{takahashi1}.

Previous high-pressure resistivity measurements on LaO$_{0.86}$F$_{0.14}$FeAs,
where pressure was always changed at RT, revealed that $T_{c}$ from the
resistivity onset increases rapidly with pressure in Fluorinert pressure
medium at the rate $+12$ K/GPa \cite{takahashi1}. At a pressure of 0.66 GPa,
therefore, $T_{c}$ would be expected to increase by approximately 8 K. In
Fig.~2(a), however, the application of 0.66 GPa He-gas pressure at RT is seen
to slightly broaden the transition in the ac susceptibility and shift it only
slightly ($\sim$ 0.2 K) to higher temperatures (point 2 in Fig.~2(b)), a shift
40$\times$ less than the 8 K expected! This difference in $dT_{c}/dP$
decreases to 15 $\times$ if the resistivity zero point is used (4.4 K/GPa).

The sample was then allowed to remain at RT for various cumulatitive lengths
of time (20 h for point 3, 68 h for point 4, 112 h for point 5) for a total of
112 h, during which time the pressure at RT decreased only slightly to 0.65
GPa; surprisingly, $T_{c}$ is seen in Fig.~2(b) to decrease by $\sim$ 0.3 K
(point 5)! Releasing the pressure then at 55 K to 0 GPa \ results in $T_{c}$
shifting further downwards to a temperature (point 6) $\sim$ 0.5 K less than
its initial value at ambient pressure (point 1), the transition recovering its
original sharpness. After the release of pressure at 55 K (point 6), holding
the sample at 270 K for 1 h did not result in a further change in $T_{c}$
(point 7). The superconducting transition appears to be stuck at this lower
value. Such a feature was not observed in oxygen ordering phenomena in the
cuprates \cite{schilling1}.

A second sample from the same synthesis batch was then studied to check these
highly anomalous results, yielding the data points (open circles) labeled with
primed numbers in Fig.~2(b). The value of $T_{c}$ at ambient pressure was
identical to that of the previous sample. Applying 0.78 GPa He-gas pressure at
RT shifted $T_{c}$ upward by only 0.23 K (point 2$^{\prime}$), yielding a
slope $dT_{c}/dP\simeq+0.30(1)$ K/GPa in excellent agreement with the results
for the first sample (closed circles), but far less (40$\times$) than that
(dashed line) observed in resistivity studies by Takahashi \textit{et al.}
\cite{takahashi1} using Fluorinert pressure medium. The pressure was then
reduced successively at low temperatures to $P=0$ ($2^{\prime}\rightarrow
3^{\prime}$ at 60 K, $3^{\prime}\rightarrow4^{\prime}$ at 55 K, $4^{\prime
}\rightarrow5^{\prime}$ at 40 K, and $5^{\prime}\rightarrow6^{\prime}$ at 35
K). At ambient pressure $T_{c}$ now lies 0.62 K lower than the initial value
(point 1$^{\prime}$). Holding the sample for 1 h at 100 K caused no further
change in $T_{c}$ (point 7$^{\prime}$). $T_{c}$ was observed to shift upwards
by 0.33 K after holding at 200 K for 1.3 h (point 8$^{\prime}$), but no
further shift in $T_{c}$ occurred after holding at 250 K for 1 h (point
9$^{\prime}$) or at RT for 30 h (point 10$^{\prime}$). A pressure of 0.33 GPa
was then applied at 60 K (point 11$^{\prime}$) and released again at 50 K
(point 12$^{\prime}$), yielding a value of $T_{c}$ approximately 0.8 K lower
than the initial value (point 1$^{\prime}$). The ambient pressure value of
$T_{c}$ did not change further, even after holding at RT for 170 h (point
13$^{\prime}$)!

\section{Discussion}

In all previous high-pressure studies on the superconducting pnictides,
pressure was changed at RT. We first compare the results of those studies to
the present He-gas results for pressure change at RT. To our knowledge, the
only measurements of $T_{c}(P)$ under pressure on LaO$_{1-x}$F$_{x}$FeAs for
fluorine concentrations near those ($x=0.07$ and 0.14) used in the present
study are the resistivity measurements to 1.5 GPa with Fluorinert pressure
medium by Takahashi \textit{et al.} \cite{takahashi1} for $x=$ 0.05, 0.08,
0.11, and 0.14. As discussed above, if the pressure is changed at RT, the
values obtained for $dT_{c}/dP$ from resistivity studies depend sensitively on
the criterion used to determine $T_{c}$; the resistivity onset or
$\rho\rightarrow0$ point give $dT_{c}/dP\simeq$ 2 or 0.54 K/GPa, respectively,
in contrast to the present ac susceptibility studies using He-gas pressure
where the intermediate value $dT_{c}/dP\simeq1.2$ K/GPa is observed. The
difference in the value of $dT_{c}/dP$\ is much larger at the higher fluorine
concentration $x=0.14,$ where Takahashi \textit{et al.} \cite{takahashi1} find
$dT_{c}/dP\simeq$ 12 K/GPa (onset) or 4.4 K/GPa ($\rho\rightarrow0$ point),
the respective values being 40$\times$ or 15$\times$ higher than the 0.30
K/GPa found in the present He-gas experiments (see Fig.~2(b)). The fact that
the anomalous temperature/pressure effects are most dramatic for the
$x=0.14$\ sample, which lies near the substitution limit of F for O, suggests
that the application of pressure may cause an irreversible phase separation.
This would explain why the value of $T_{c}$ does not recover in our
experiments after a pressure cycle, as seen in Fig.~2(b)\ 

We note that at the lower concentration $x=0.11$ Takahashi \textit{et al.}
\cite{takahashi1} report $dT_{c}/dP\simeq$ 8 K/GPa. On the other hand, at the
same F-concentration the much lower value $dT_{c}/dP\approx1.2$ K/GPa is
obtained in a dc susceptibility measurement by Lu \textit{et al.} \cite{lu} to
1 GPa pressure using an unspecified fluid pressure medium and in resistivity
(onset) studies by Zocco \textit{et al}. \cite{zocco} using \textit{n}-pentane
: iso-amyl alcohol pressure medium to 0.94 GPa. It appears, therefore, that in
the 1111 Fe-pnictides the pressure dependence of $T_{c}$ depends sensitively
not only on the dopant concentration but also on which physical property is
measured, how the value of $T_{c}$ is determined, and the type of pressure
transmitting medium used. This would appear to support the view of Yu
\textit{et al. }\cite{yu} that shear stress effects play an important role in
determining $T_{c}(P)$ in the oxypnictides, large shear stresses generating
significant changes in $T_{c}$. The marked temperature/pressure history
effects seen in Figs.~1(b) and 2(b) may be indicative of important
shear-stress effects between grains in polycrystalline samples, even when
purely hydrostatic He-gas pressure is applied. Parallel measurements on high
quality single crystals would test this hypothesis. It is also possible that
short-range diffusion of oxygen or flourine anions within the crystal
structure may occur in response to a change in pressure at RT, much as the
\textquotedblleft oxygen ordering effects\textquotedblright\ observed in the
cuprate oxides \cite{schilling1}.

In order to check whether the temperature/pressure history effects seen here
might result from the penetration of the He pressure medium into the crystal
lattice, we heated a 3.3 mg portion of the $x=0.14$ sample used in the present
He-gas experiments to 100$^{\circ}$C while connected to a sensitive mass
spectrometer. We were unable to detect the slightest trace of He escaping from
the sample. Subsequent vaporization of this sample in an ultra-sensitive mass
spectrometer set the He impurity level at $\sim$ 1 ppm, an amount far too
small to effect the dramatic changes observed in $T_{c}$.

An alternative scenario is conceivable. The undoped compound LaOFeAs exhibits
a spin-density-wave (SDW) and structural phase transition
(tetragonal$\rightarrow$orthorhombic) below 150 K \cite{kamihara,cruz}.
Substituting O with F or applying pressure are believed to suppress this
transition and allow a superconducting ground state to appear. A competition
between a SDW instability and superconductivity is also observed in
CeO$_{1-x}$F$_{x}$FeAs \cite{chen}. Perhaps shear-stress effects result in
superconducting and nonsuperconducting SDW regions in the sample which lead to
the complex temperature/pressure history effects found in the present studies.

Whatever the explanation for the anomalous behavior of $T_{c}(P)$, it is
likely that in the oxypnictides, as in the cuprates, a full understanding of
the manner in which $T_{c}$ changes under pressure may be difficult to obtain
since it almost certainly depends on several factors simultaneously, including
the strength of shear-stress effects as well as changes in the carrier
concentration and the separation and area of the superconducting planes.
Further experimentation is clearly needed here which focusses both on
pressure-induced changes in the superconductivity and crystal structure on
global and local scales.

\vspace{0.3cm}

\noindent\textbf{Acknowledgments.} The authors grateful to A. Meshik at
Washington University for carrying out the ultra-sensitive mass spectrometry
measurements. The high-pressure studies at Washington University were
supported by the National Science Foundation through Grant No. DMR-0703896.

\begin{center}
\bigskip{\LARGE Figure Captions}
\end{center}

\bigskip\ 

\noindent\textbf{Fig. 1. \ }LaO$_{0.93}$F$_{0.07}$FeAs sample (mass 1.54 mg).
(a) Real part of the ac susceptibility versus temperature at four selected
He-gas pressures. Order of measurement given by pressure values from top to
bottom. Large cross (+) marks midpoint of superconducting transition for
initial measurement at ambient pressure (0 GPa). (b) $T_{c}$ versus pressure
for all data taken. Numbers give order of measurement (see text for details).
Closed and open circles used for clarity. Solid straight line is guide to eye;
dashed line gives slope inferred from resistivity onset data in Ref.
\cite{takahashi1}.\bigskip

\noindent\textbf{Fig.\ 2.\ \ }LaO$_{0.86}$F$_{0.14}$FeAs sample. (a) Real part
of the ac susceptibility versus temperature at four selected He-gas pressures
for sample with mass 8.69 mg. Order of measurement given by pressure values
from top to bottom (corresponds to points 1, 2, 5, 6 in Fig. 2(b)). (b)
$T_{c}$ versus pressure for all data taken. Numbers give order of measurement
(see text for details). Data for primed and unprimed numbers (open and closed
circles) taken on different samples with masses 5.65 mg and 8.69 mg,
respectively. Solid straight line is guide to eye; dashed line gives slope
from resistivity onset data in Ref. \cite{takahashi1}.
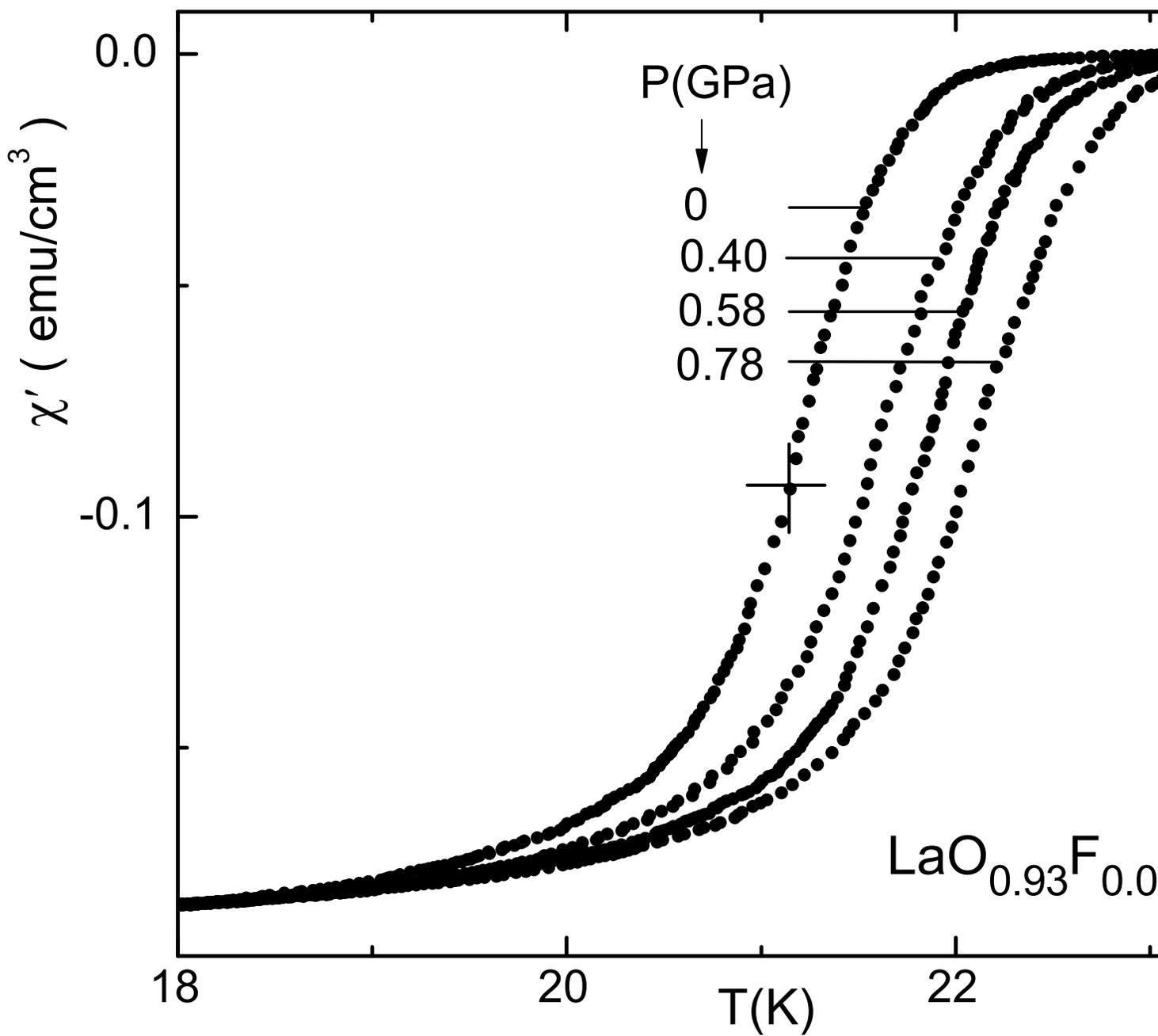
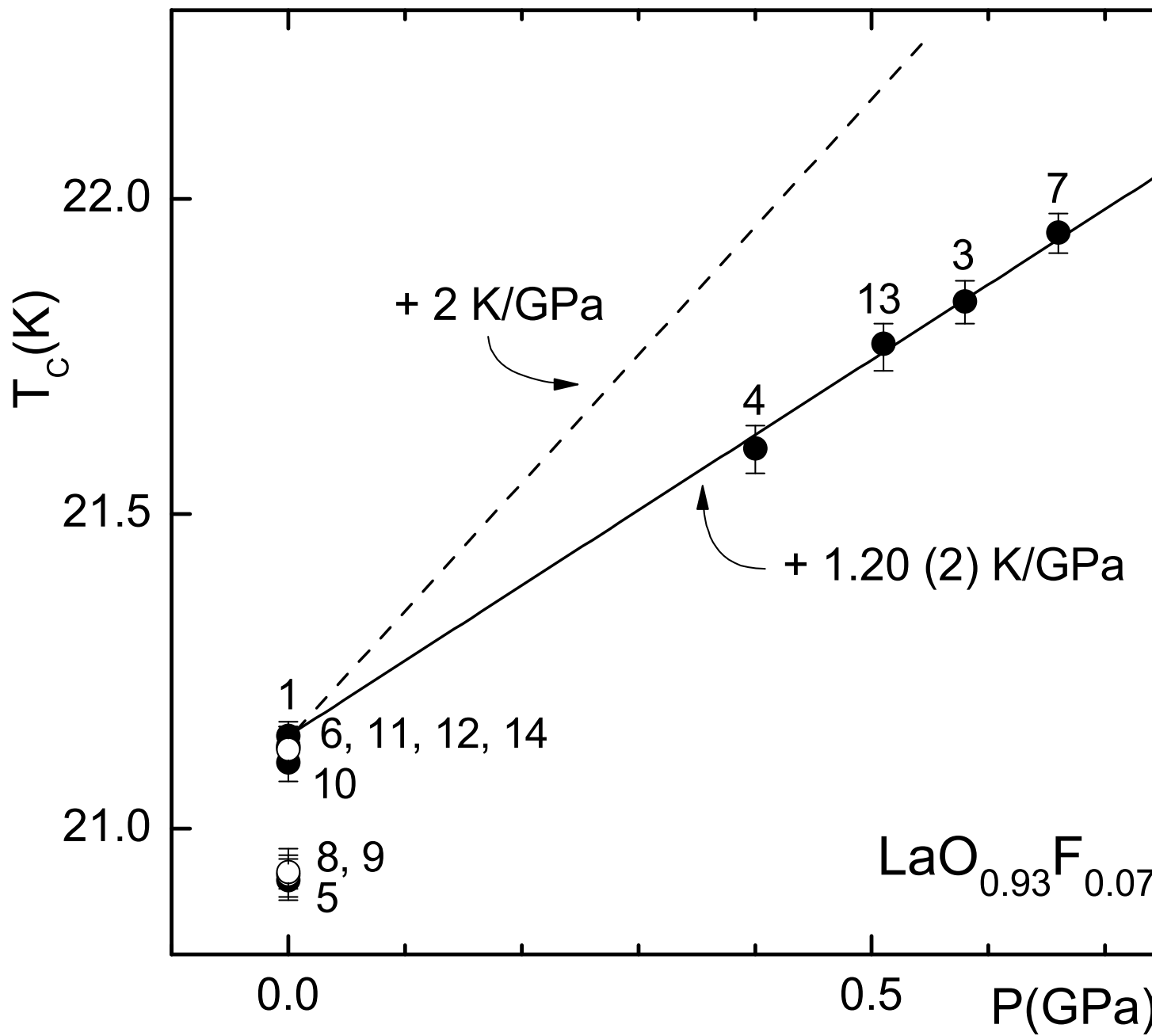
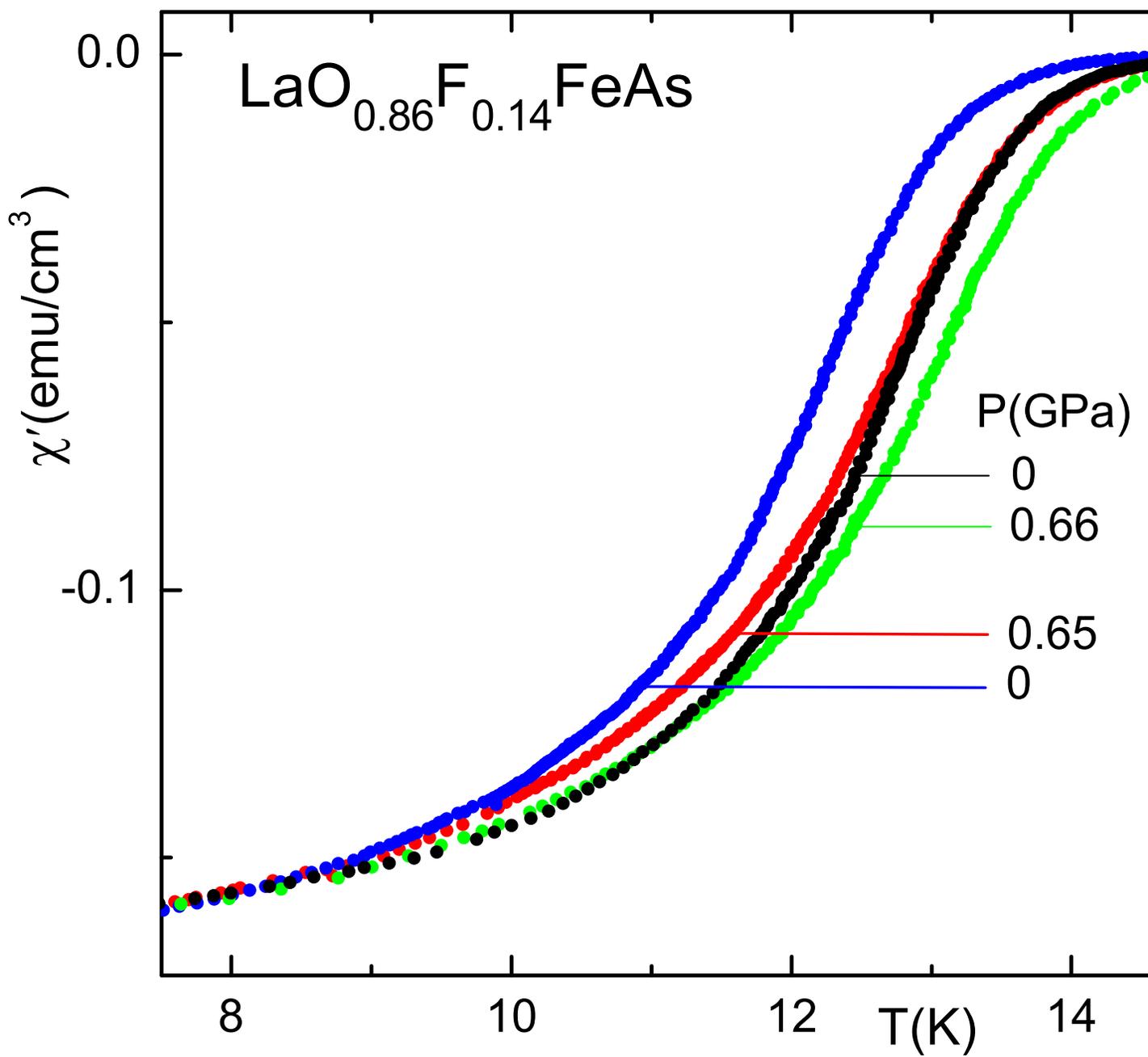
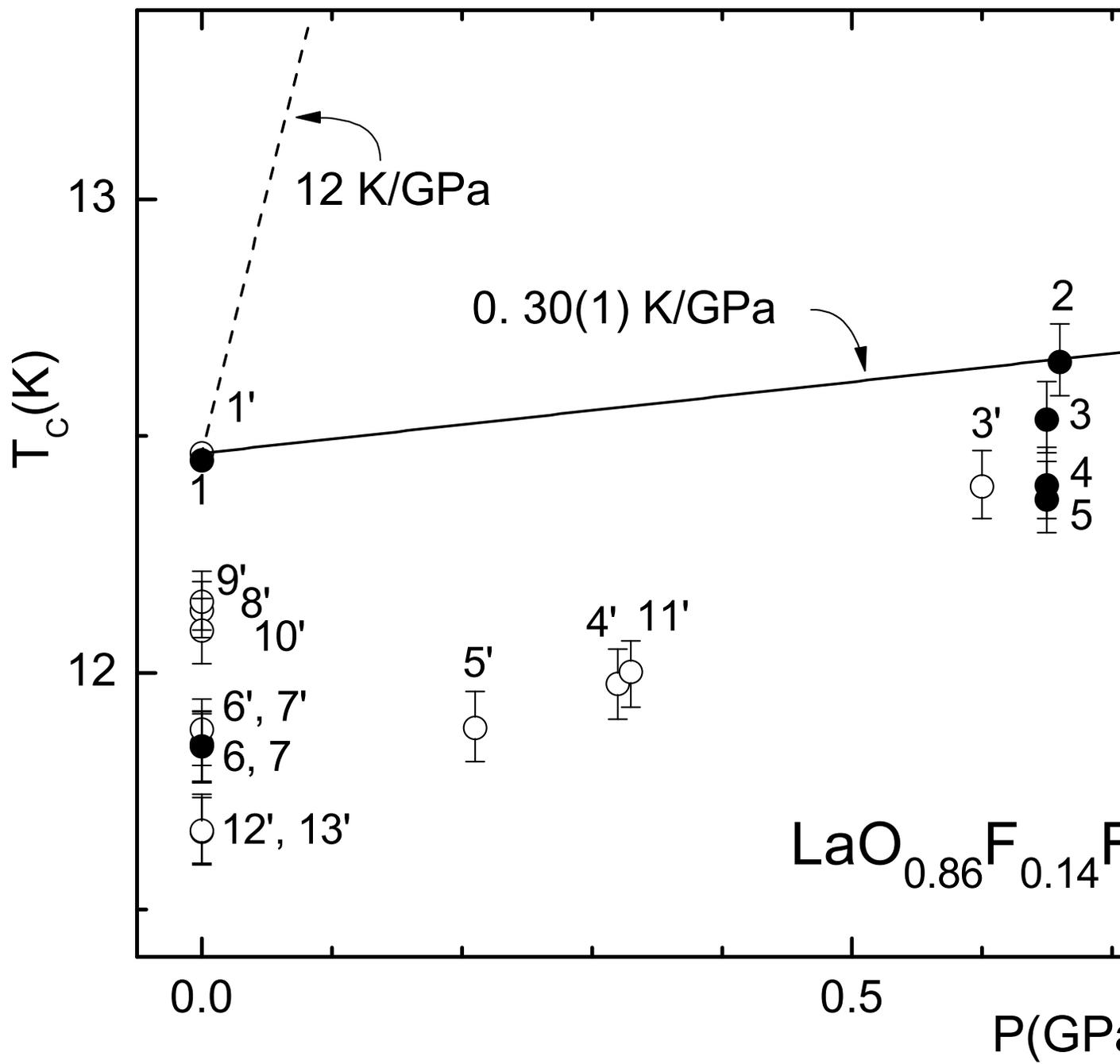
\end{document}